\documentstyle[12pt,epsf]{iopuot}          
\begin{document}
\jjll{1} 
\title{Extrapolation-CAM Theory for Critical Exponents}[Extrapolation-CAM
       Theory]

\author{Howard L Richards\dag\ftnote{2}
   {Present address: Max-Planck-Institut f{\" u}r Polymerforschung, 
    D-55128 Mainz, Germany},
  Naomichi Hatano\dag\ftnote{3}{Present address:
        Theoretical Division, 
        Los Alamos National Laboratory, Los Alamos, NM~87545, USA},
and M~A~Novotny$\parallel$ \P 
}

\address{\dag\ Department of Physics, University of Tokyo, 
         Tokyo 113, Japan}
\address{$\parallel$\ 
  Supercomputer Computations Research Institute,  \\
  Florida State University, Tallahassee, Florida 32306-4052, USA
        }
\address{\P\ Department of Electrical Engineering, \\
  2525 Pottsdamer Street, 
  Florida A\&M University--Florida State University, \\
  Tallahassee, Florida 32310-6046, USA
       }

\begin{abstract}
We propose and test a new method for generating canonical 
sequences for analysis by the Coherent Anomaly Method (CAM) 
{}from non-mean-field approximations. 
By intentionally underestimating the rate of convergence of
exact-diagonalization values for the mass or energy gaps 
of finite systems, we form families of sequences of gap estimates.
The gap estimates cross zero with generically nonzero linear 
terms in their Taylor expansions, so that $\nu \! = \! 1$ for 
each member of these sequences of estimates.  Thus, the 
CAM can be used to determine $\nu$. 
Our freedom in deciding exactly how to underestimate the 
convergence allows us to choose the sequence that displays 
the clearest coherent anomaly. 
We demonstrate this approach on the two-dimensional ferromagnetic 
Ising model, for which $\nu \! = \! 1$. 
We also use it  on the three-dimensional 
ferromagnetic Ising model, finding 
$\nu \! \approx \! 0.629$, in good agreement with other 
estimates. 
Finally, we apply it to an antiferromagnetic spin-1 Heisenberg 
chain, finding $\nu \! \approx \! 0.987$ at the phase transition 
between the Haldane phase and the dimerized phase, in agreement 
with the field-theoretic prediction $\nu \! = \! 1$. 
Although the specific systems used to test the extrapolation-CAM 
procedure involve finite system sizes, the method could be 
applied to other finite approximations, such as systematic 
variational approximations. 
\end{abstract}


\section{Introduction} 
\label{sec-intro} 

On approaching a critical point, some quantity diverges in the 
thermodynamic limit with a characteristic critical exponent. 
The Coherent Anomaly Method (CAM) has proven quite successful 
in determining critical exponents from certain sequences of 
approximations~\cite{SuzukiCAM,SuzukiCAF,SuzukiTACVPPM}.
The CAM requires a systematic, or {\em canonical}, sequences 
of approximations, all of which yield identical, known critical 
exponents.  The prototypical example is a sequence of mean-field 
approximations in which successively larger clusters allow 
more and more fluctuations to be properly taken into 
account~\cite{SuzukiCam1}; all of the critical exponents 
in this example assume their ``classical'' values. 
The CAM uses the known critical exponents of the approximate 
systems and the behaviour of the critical {\em amplitudes} 
as the quality of approximation is improved to determine 
the true critical exponents of the original system being 
approximated. 
The purpose of this paper is to show how sequences of 
approximations in which the critical points are ill 
defined and divergences do not occur can be used to 
construct canonical sequences through extrapolation. 

The basic idea is as follows.  Suppose that in the 
thermodynamic limit a nonnegative 
quantity $\xi(\beta)$ diverges as a system parameter 
$\beta$ approaches a critical value 
$\beta_{\rm c}^*$ as 
\begin{equation}
   \label{eq:defnuxi}
   \xi (\beta) \sim (\beta_{\rm c}^* - \beta)^{-\nu} \; ,
\end  {equation} 
where $\nu$ is a critical exponent of unknown value. 
The reciprocal of $\xi(\beta)$ then converges to zero as 
\begin{equation}
   \label{eq:defnu}
   \Delta (\beta) \equiv [\xi(\beta)]^{-1}
    \sim (\beta_{\rm c}^* - \beta)^\nu \; . 
\end  {equation} 
Suppose further that we have a sequence of monotonically 
decreasing approximations $\{\delta_i(\beta)\}$ such that 
\begin{equation}
  \label{eq:condit} 
  \delta_i(\beta)  >  \delta_{i+1}(\beta)  
  > \delta_\infty (\beta)  \equiv  \Delta(\beta) 
  \qquad \forall(i,\beta) \; .  
\end{equation}
By taking two or more consecutive 
values of $\delta_i (\beta)$, we make an extrapolation 
$\Delta_{\{i\}} (\beta)$ which {\em intentionally underestimates} 
the rate of convergence in $\{\delta_i (\beta)\}$.
Clearly, these extrapolations will be negative at 
$\beta_{\rm c}^*$, and they generically cross zero as 
\begin{equation}
  \label{eq:exCAM} 
  \Delta_{\{i\}}(\beta) \simeq 
         \left( -\left.
         \frac{\d \Delta_{ \{i\} } } {\d \beta}
         \right|_{\tsty \beta = \beta_{ \{i\} ,{\rm c}}} \right)
         (\beta_{ \{i\} ,{\rm c}} - \beta)      
\end  {equation}  
for some $\{\beta_{ \{i\},{\rm c}}\}$. For any 
reasonable extrapolation procedure (e.g.\ power-law 
extrapolation or exponential extrapolation),  
$\lim_{ \{i\} \rightarrow \infty} \Delta_{\{i\}} (\beta)
  \! = \! \Delta(\beta)$.   
The CAM hypothesis, which can be justified on the 
basis of an envelope argument~\cite{SuzukiCam1}, is that 
\begin{equation}
   \label{eq:hypo}
   -\left.\frac{\d \Delta_{\{i\}} }{\d \beta}
         \right|_{\tsty \beta = \beta_{ \{i\} ,{\rm c}}} 
         \sim (\beta_{\rm c}^* - \beta_{ \{i\} ,{\rm c}})^{\nu - 1}\; ,
\end  {equation}
so that, in analogy with \eref{eq:defnu}, 
\begin{equation}
  \label{eq:anlgy} 
   \Delta_{\{i\}}(\beta) \sim 
   (\beta_{\rm c}^* - \beta_{ \{i\} ,{\rm c}})^{\nu - 1}
         (\beta_{ \{i\} ,{\rm c}} - \beta)   \; . 
\end  {equation}
This provides us with a convenient means of measuring $\nu$. 
According to \eref{eq:hypo}, a plot of 
\begin{equation}
  \label{eq:defY} 
    Y_{ \{i\} } \equiv \ln \left( -\beta_{ \{i\} ,{\rm c}} 
                      \left. \frac{\d \Delta_{  \{i\} }} {\d \beta} 
                   \right|_{\tsty \beta = \beta_{ \{i\} ,{\rm c}} }  
                     \right)
\end  {equation}
versus 
\begin{equation}
  \label{eq:defX} 
  X_{ \{i\} } \equiv \ln \left(1 - 
          \frac{\beta_{ \{i\} ,{\rm c}} }{\beta_{\rm c}^*}\right)
\end  {equation}
should (in the limit $X \! \rightarrow \! -\infty$) 
be a straight line with slope $\nu \! - \! 1$.
Furthermore, since all that is required of the extrapolation is 
that it must  underestimate the convergence, we are free to choose 
an extrapolation which leads to a particularly clear coherent 
anomaly. 

The organization of the remainder of this paper is as follows. 
In \sref{sec-2DI} we take as $\Delta(\beta)$ the mass gap, i.e.,  
the reciprocal of the correlation length, 
as a function of inverse temperature in the square-lattice Ising 
ferromagnet. This model has the advantage that 
the correlation length is known analytically~\cite{Onsager44}. 
In \sref{sec-FSS} we analytically study the asymptotic behaviour 
of the estimated critical exponent if $\delta_{i}$ is given 
by a finite-size scaling function. 
In \sref{sec-3DI} we study the critical behaviour of the 
mass gap in the cubic-lattice Ising ferromagnet, and demonstrate 
that the method works even when $\nu \! \neq \! 1$. 
In \sref{sec-1DH} we study the critical behaviour of the 
energy gap in an antiferromagnetic spin-1 Heisenberg chain 
with bilinear and biquadratic interactions at the phase transition 
between the Haldane and dimerized phases. 
In \sref{sec-concl} we summarize and discuss possible extensions 
of this work.

Note that although all of the initial approximations $\delta_{i}$ 
used in this study come from systems that are finite in 
at least one dimension, this is only to 
provide convenient examples.  The method itself does not restrict 
us to such systems. 

\section{The Square-Lattice Ising Ferromagnet}
\label{sec-2DI} 

As the first example, we use the extrapolation-CAM method to 
determine $\nu$ for the 
classical Ising ferromagnet with Hamiltonian 
\begin{equation}
  \label{eq:HIsing}
  {\cal H} = -\sum_{\langle i,j \rangle} s_i s_j \; ,
\end  {equation}
where $s_i \! = \! \pm 1$.  The sum $\sum_{\langle i,j \rangle}$ 
runs over all  nearest-neighbour pairs on a periodic square lattice 
which is of length $L$ 
in the $y$-direction and of infinite length 
in the $x$-direction.  
The unit of length is the lattice constant. 

This model has some very advantageous properties:
the mass gaps $\{\delta_L(\beta)\}$ can be 
calculated analytically for systems of arbitrary finite width $L$, 
and the mass gap $\Delta(\beta)$ for the thermodynamic limit of 
the model can also be calculated analytically~\cite{Onsager44}. 
Consequently, we can make a detailed comparison of the 
extrapolation-CAM estimates for the critical exponent $\nu$
with its rigourously known value, $\nu \! = \! 1$. 

\begin{figure}
      \epsfxsize=\textwidth
      \epsfbox{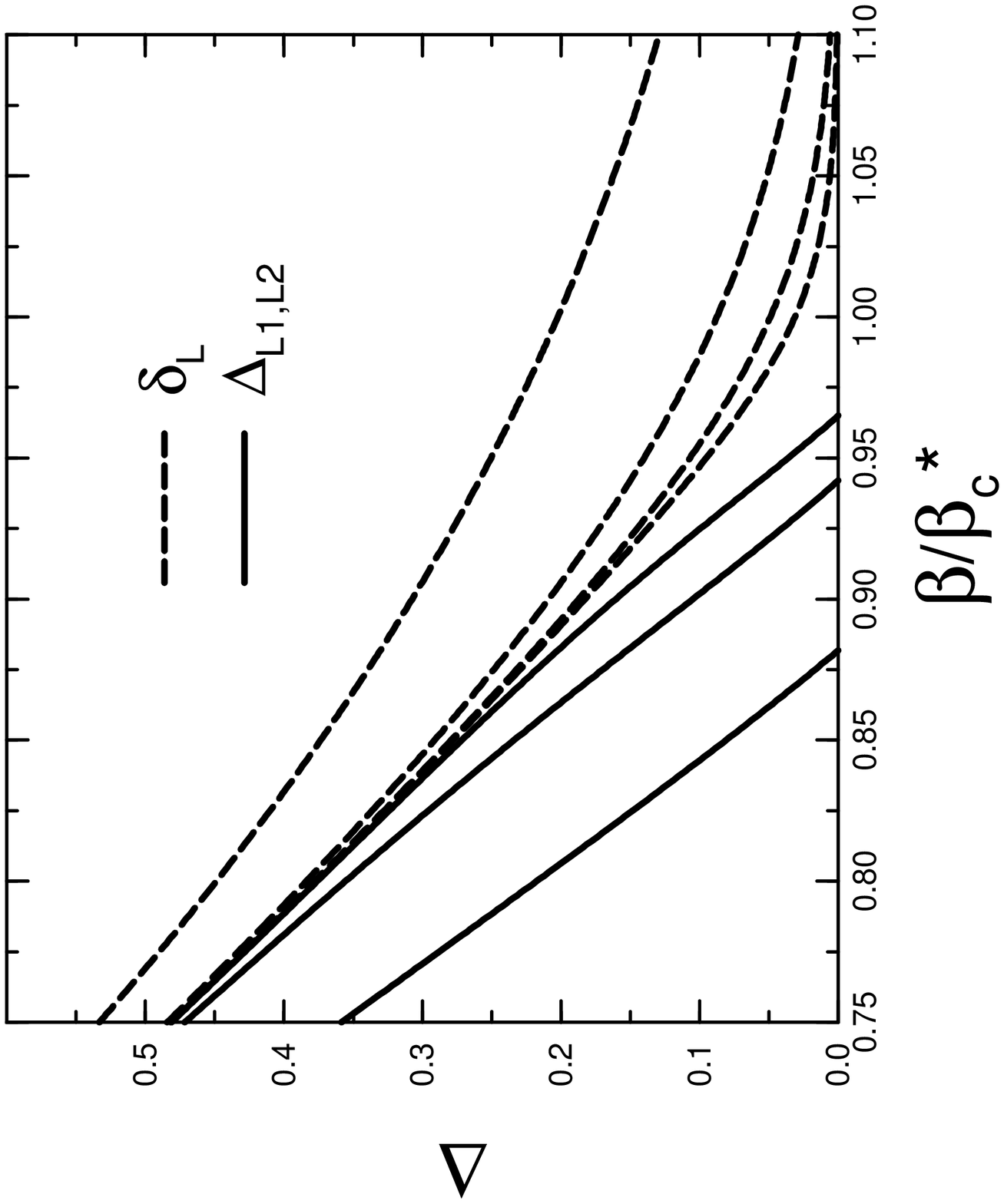}
          \noindent 
        \caption
                { 
                 \label{f:2DGAP}
                 The mass gap, or inverse correlation length, 
                 for the square-lattice Ising ferromagnet. 
                 The dashed lines are the gaps for semi-infinite
                 systems of width 4, 9, 16, and 25, from top to 
                 bottom.  The solid lines are extrapolations 
                 using \protect\eref{eq:XSolve2} with 
                 the value of $B$ determined in 
                 \protect\fref{f:2Dnu}; from left to right, they 
                 represent $(L_1,L_2)$ = (4,9), (9,16), and (16,25). 
                }
\end{figure}

For this model the variable 
$\beta$ is the reciprocal of the dimensionless temperature.  
Since it is known that $\delta_L \! \sim \! L^{-1}$ 
for sufficiently large systems at the 
critical point~\cite{Fisher71,Fisher72,Barber_PTCPv8}, 
we extrapolate by solving 
\begin{equation}
  \eqalign{
  \label{eq:XSolve2} 
  \delta_{L}       (\beta) &= \Delta_{L,L^\prime}(\beta) 
             + A(\beta) L^{-B}  \\
  \delta_{L^\prime}(\beta) &= \Delta_{L,L^\prime}(\beta) 
             + A(\beta) L^{\prime-B} 
          }
\end  {equation} 
for $\Delta_{L,L^\prime}(\beta)$ at fixed $\beta$ for 
$B \! \leq \! 1$. 
The result of such an extrapolation is shown in 
\fref{f:2DGAP}.
  
\begin{figure}
      \epsfxsize=\textwidth
      \epsfbox{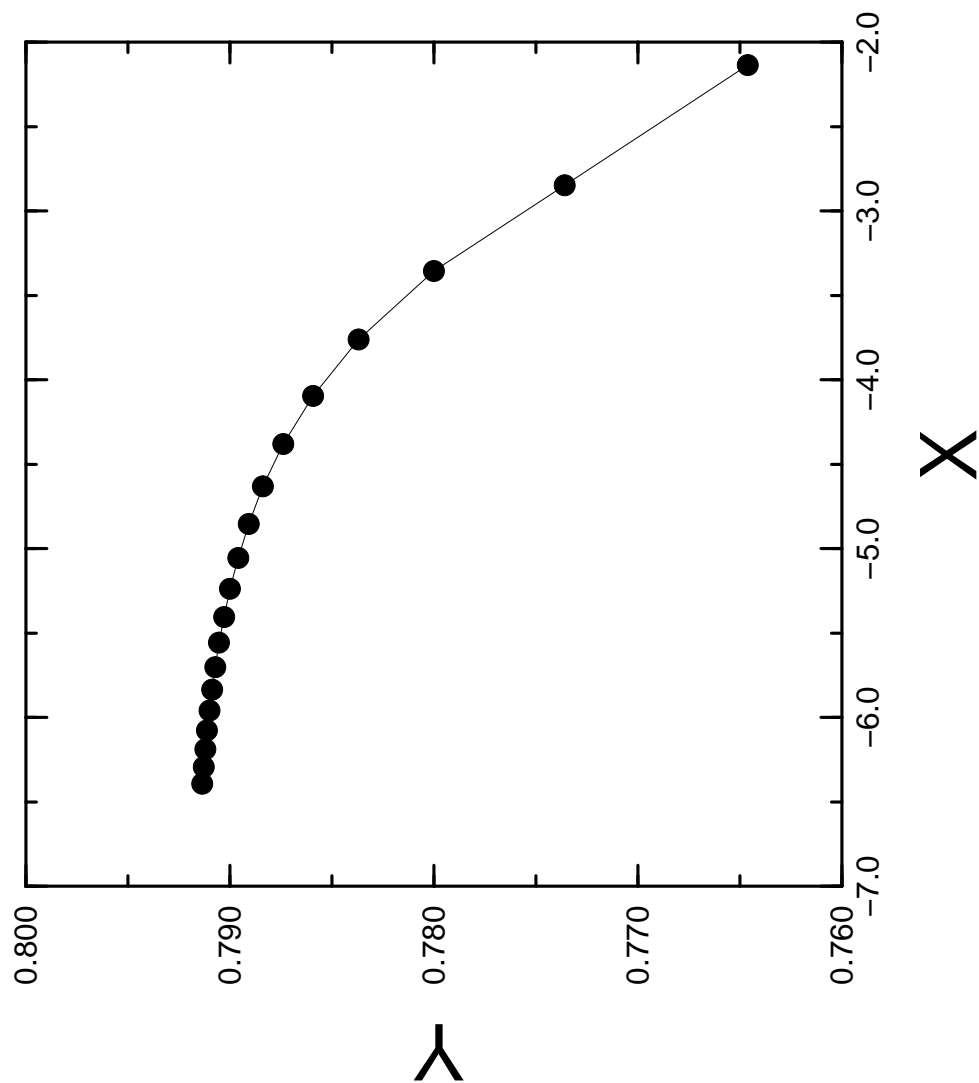}
          \noindent 
        \caption
                { 
                 \label{f:2DCAM}
                 CAM plot for the square-lattice Ising model.
                 Equations~(\protect\ref{eq:defX}) and 
                 (\protect\ref{eq:defY}) define $X$ and $Y$. 
                 Extrapolations use the value of $B$ determined in 
                 \protect\fref{f:2Dnu}                 
                 and system sizes that are consecutive perfect squares: 
                 $L \in \{4, 9, 16, \ldots, 400\}$. 
                 Note the difference in the scales.  
                }
\end{figure}

\begin{figure}
      \epsfxsize=\textwidth
      \epsfbox{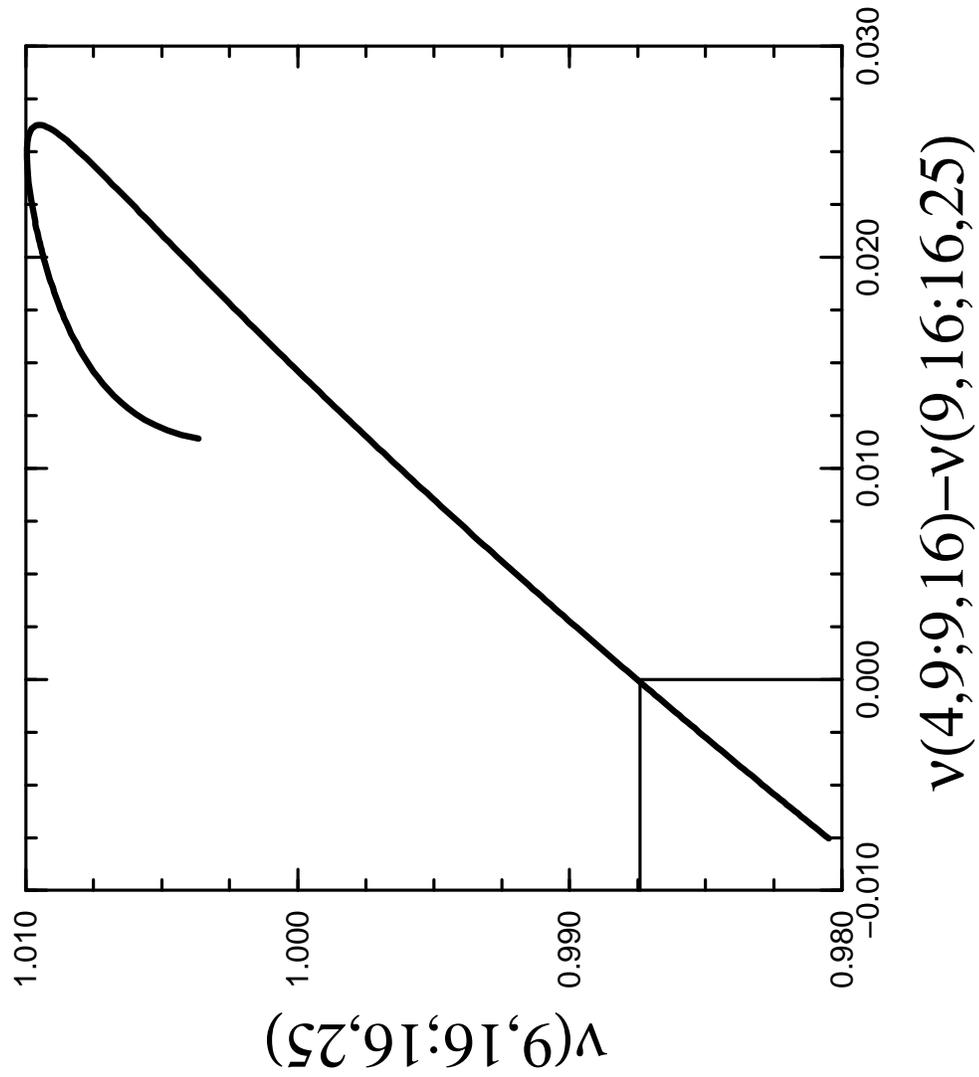}
          \noindent 
        \caption
                { 
                 \label{f:2Dnu}
                 By varying $B$, $\nu$ can be estimated from 
                 consecutive extrapolations.  In order to 
                 determine a good estimate of $\nu$, we vary 
                 $B$ until two consecutive estimates of $\nu$ 
                 are identical.  This yields  
                 $B \! = \! 0.407\,404\,833$ and  
                 $\nu \! = \! 0.987\,405\,623$. 
                }
\end{figure}

\begin{figure}
      \epsfxsize=\textwidth
      \epsfbox{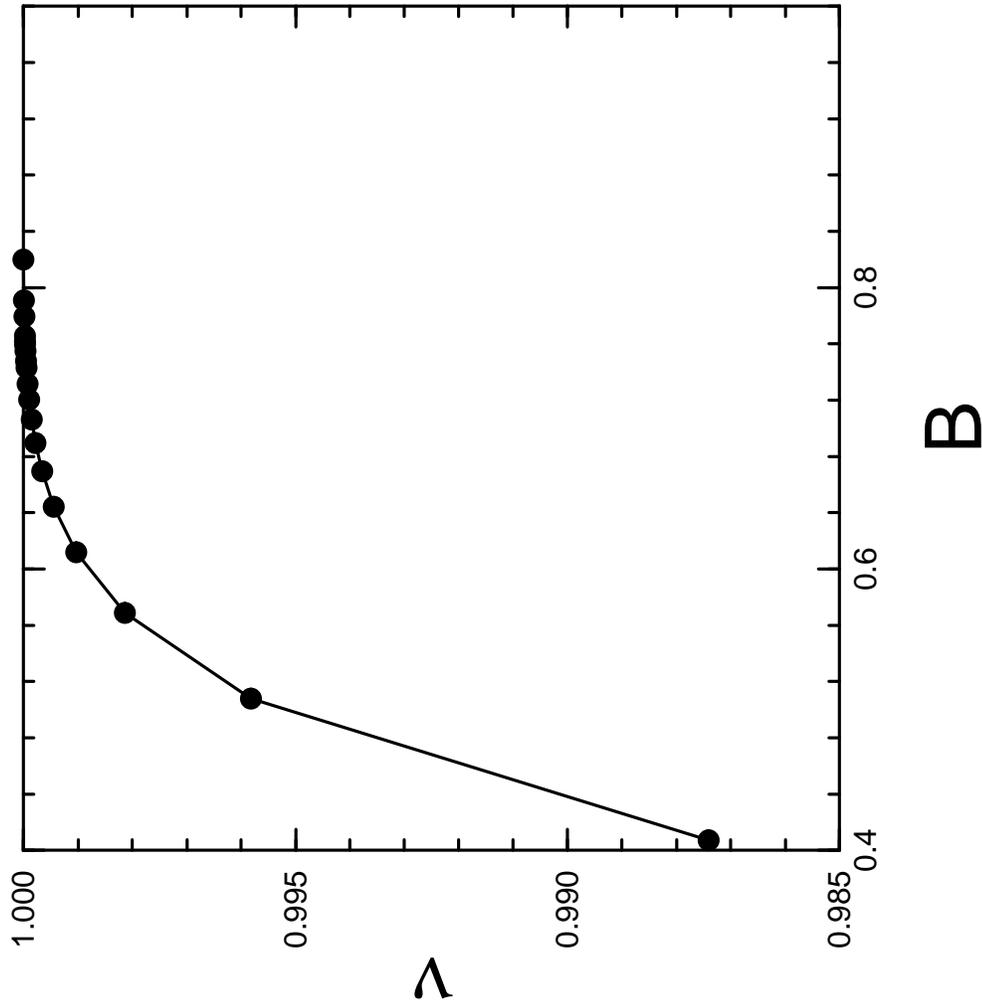}
          \noindent 
        \caption
                { 
                 \label{f:2DnuB}
                 Using consecutive perfect squares, i.e.\ 
                 $L_j \! = \! (k + j -1)^2$, 
                 $j \in \{ 1, 2, 3, 4\}$ and 
                 $k \in \{ 1, 2, 3, \ldots\}$, we determine 
                 $\nu$ and $B$ as in \protect\fref{f:2Dnu}. 
                 As the system sizes become large, it appears 
                 that $\nu \! \rightarrow \! 1$ and 
                 $B \! \rightarrow \! 1$.  For some other 
                 series $\{L_j\}$, the convergence is much 
                 less clear. 
                }
\end{figure}

\Fref{f:2DCAM} shows $Y$ vs.\ $X$, defined by \eref{eq:defY} and 
\eref{eq:defX}, to be a curve with 
a small slope. The slope tends to zero as $-X$ becomes large --- 
that is, when $\beta_{L,L', \rm c}$ becomes a very good 
approximation to $\beta_{\rm c}^*$. 
We can estimate $\nu$ from the slope of the line connecting 
two adjacent points $(X_1,Y_1)$ and $(X_2,Y_2)$. This 
estimate will depend not only on the quality of the initial 
approximations $\delta_L(\beta)$, 
but also on the parameter $B$.  We can 
constrain $B$ by taking a third point $(X_3,Y_3)$ and demanding 
that the three points be colinear, so that the value of 
$\nu_{\{L\}}$ will be unambiguous. 
\Fref{f:2Dnu} shows this for values of $\nu$ based on 
systems with 
$L \! = \! 4$, 9, 16, and 25.  The resulting estimate is
$\nu_{\{L\}} \! = \! 0.987\,406$, in good agreement with the 
exact value $\nu \! = \! 1$.  
As may be expected, increasing the sizes of the four 
systems needed to form the estimate of $\nu$ increases 
the accuracy of $\nu_{\{L\}}$.
\Fref{f:2DnuB} shows the estimated value of $\nu$ 
vs.\ $B$ for  $L \! = \! j^2$, $(j+1)^2$, $(j+2)^2$, and $(j+3)^2$, 
where $j \! \in \! \{2,3,4, \ldots, 20\}$. 
The estimate for $j \! = \! 20$ is $\nu \! = \! 1.000\,004$. 
Further increase of $j$ actually causes the accuracy 
of the estimate to become worse due to the increasingly 
large sums required to calculate $\delta_L (\beta)$ and 
$\d \delta_L / \d \beta$ and the finite 
(eight-byte) numerical precision of our programs. 
The convergence of $\nu_{\{L\}}$ depends on how 
the four system sizes are chosen, and can be both 
complicated and nonmonotonic. 

\section{Relation to Finite-Size Scaling}
\label{sec-FSS} 

Some insight into the dependence of $\nu_{\{L\}}$ on 
the system sizes can be gained by 
assuming that $\delta_{L}$ satisfies the scaling 
equation~\cite{Barber_PTCPv8}
\begin{equation}
  \label{eq:scale}
  \delta_{L}(\beta) = L^{\omega}Q(x,y) \, , 
\end  {equation}
where 
\begin{equation}
  \label{eq:smallx}
  x \equiv \left( 1 - \frac{\beta}{\beta_{c}^{*}}  \right) L^{\theta}
\end  {equation}
and 
\begin{equation}
  \label{eq:smallz}
  y \equiv \zeta L^{\phi} \, .
\end  {equation}
This assumption is well justified for the systems actually analyzed in 
this paper, since they all are finite in at least one dimension. 
In order to be consistent with \eref{eq:defnu}, we must have 
$\omega/\theta \! = \! -\nu$, and 
$\theta \! = \! y_{T}$~\cite{Barber_PTCPv8}. 
The variable $\zeta$ is the correction-to-scaling amplitude 
and $\phi$ is the correction-to-scaling exponent, where 
the correction to scaling is assumed to arise from the leading 
irrelevant field~\cite{Wegner72}.

For large $L^{\prime} \! = \! L \! + \! 1$, \eref{eq:XSolve2} 
yields 
\begin{eqnarray}
  \Delta_{L,L^\prime}(\beta) & = & \delta_{L}(\beta) 
    - \frac{\delta_{L}(\beta) - \delta_{L^{\prime}}(\beta)}
           {L^{-B} - L^{\prime-B}} L^{-B} \nonumber \\
 & = & B^{-1}L^{\omega} 
        \Biggl[ (B + \omega)Q(x,y) 
        + \theta x \frac{\partial Q}{\partial x}(x,y) 
\nonumber \\ & & \label{eq:scD_Nb}  
        + \phi   y \frac{\partial Q}{\partial y}(x,y) 
         + O(L^{-1}) \Biggr] \, . 
\end  {eqnarray}
For the remainder of this section, we consider $L$ to be a 
single continuous variable and drop the second index $L'$.
The $O(L^{-1})$ term in \eref{eq:scD_Nb} comes from the 
truncation error in a difference formula approximation 
of a derivative.  This term cannot be neglected if 
$\phi \! \leq \! -1$, unless we generalize the 
extrapolation procedure to allow for a sufficiently 
high-order difference formula~\cite{BurdenFaires4.1};
for the remainder of this section we assume that this 
is done if necessary and neglect the truncation error. 

For the moment, we set $\zeta \! = \! 0$ and perform 
the extrapolation-CAM procedure in the absence of 
corrections to scaling. 

It is convenient at this point to use a slightly different 
definition for $Y$, 
\begin{equation}
  \label{eq:defYb}
  Y \equiv \ln \left(-\beta_{\rm c}^{*} 
                \left.\frac{\d \Delta_{L}}{\d \beta}\right|_{L}
               \right) \, . 
\end  {equation}
Using \eref{eq:scD_Nb}, we find 
\begin{eqnarray}
  \label{eq:scY}
  Y & = & \ln \left[x\left(1-\frac{\beta}{\beta_{\rm c}^{*}}\right)^{-1}  
          \left.\frac{\d \Delta_{L}}{\d x}\right|_{L}
                \right]
\nonumber \\ & = & \left(1 + \frac{\omega}{\theta}\right)
                   \left[\ln x - X \right] - \ln B 
    + \ln \frac{\partial F}{\partial x}(x,B) \, .
\end  {eqnarray}

We find $\beta_{L,{\rm c}}$ from the condition 
$\Delta_{L}(\beta_{L,{\rm c}}) \! = \! 0$, which implies 
\begin{equation}
  \label{eq:defF}
  F(x,B) \equiv (B + \omega)Q(x,0) 
        + \theta x \frac{\partial Q}{\partial x}(x,0) = 0 \, .
\end  {equation}
The estimate for $\nu$ is given by calculating $\d Y/\d X$ 
while holding both $F(x,B)$ and $B$ constant, 
which is accomplished by holding 
$x$ constant: 
\begin{equation}
  \label{eq:scnu}
   \nu_{L} = 1 + \left.\frac{\d Y}{\d X}\right|_{x} = 
   -\frac{\omega}{\theta} 
   \equiv \nu \, . 
\end  {equation}
Finally, the ``local straightness'' constraint 
\begin{equation}
  \label{eq:scconstr}
  \left.\frac{\d^{2}Y}{\d X^{2}}\right|_{x} =  0
\end  {equation}
is simply the identity $0 \! = \! 0$.
Thus without corrections to scaling, the $\nu_{L} \! = \! \nu$ 
but $x$ and $B$ are undetermined.  Note, however, that if 
$x \! = \! 0$, $B \! = \!-\omega \! > 0$.

In order to study the effects of corrections to scaling, 
we could expand \eref{eq:scD_Nb} and the left-hand side 
of \eref{eq:scconstr} (which is proportional to $\zeta$) 
in $x$, $y$, and $\epsilon \! \equiv \! B \! + \! \omega$; 
then the requirements $\Delta_{L}(\beta_{L,\rm c}) \! = \! 0$ 
and $\d^{2} Y /\d X^{2} \! = \! 0$
yield a pair of equations of the form 
\begin{equation}
  \eqalign{
  \label{eq:scMat} 
   M_{11} x + M_{12} \epsilon = v_{1}y \\
   M_{21} x + M_{22} \epsilon = v_{2}y 
          } \, ,
\end  {equation}
where $\mat{M}$ and $\vec{v}$ are constants.
The solution of \eref{eq:scMat},
\begin{equation}
   \label{eq:solveMat}
   \left( \begin{array}{c}
           x \\ \epsilon 
          \end  {array} \right) 
   = \mat{M}^{-1} \vec{v} y \, , 
\end  {equation}
shows that both $x$ and $\epsilon$ are proportional to $y$. 
In the same way, we can expand 
the left-hand side of \eref{eq:scnu} in terms of 
$x$, $y$, and $\epsilon$.  The result then is that 
\begin{equation}
  \label{eq:Bscaleform}
   B \simeq -\omega + O(L^{\phi})
\end  {equation}
and 
\begin{equation}
  \label{eq:nuscaleform}
   \nu_{L} \simeq \nu + O(L^{\phi}) \, . 
\end  {equation}
Privman and Fisher have shown that 
the asymptotic convergence for finite-size scaling 
renormalisation techniques is also of the form 
\eref{eq:Bscaleform}~\cite{Privman83}; their 
discussion of the difficulties in actually observing 
this asymptotic behaviour should be relevant in the 
present case as well.  
For the two-dimensional Ising model, $\omega=-1$, so 
we should expect $\lim_{L\to\infty} B \! = \! 1$, as 
indeed seems plausible from \fref{f:2DnuB}. 

\section{The Cubic-Lattice Ising Ferromagnet}
\label{sec-3DI} 

In three dimensions, the Ising model defined by 
\eref{eq:HIsing} has not been solved analytically, but it 
has been the subject of a large amount of numerical study. 
We use a  Monte Carlo Renormalisation Group estimate for the 
critical point of the cubic-lattice Ising model, 
$\beta_{\rm c}^* \! = \! 0.221\,652(4)$~\cite{MCRGTc}.
Recent estimates of $\nu$ include 
$\nu \! = \! 0.642(2)$~\cite{MCRGTc},
$\nu \! \approx \! 0.631$~\cite{Kolesik95}, and 
$\nu \! \approx \! 0.646$~\cite{JYChoi97}. 

As a result of some earlier studies~\cite{MRS0,Richards93}, 
we have transfer-matrix results already available for 
cubic-lattice Ising ferromagnets with periodic boundary 
conditions and square $L \! \times \! L$ cross-sections, 
where $L \! \in \! \{2,3,4,5\}$.  
These system sizes are obviously quite small, and are not really 
competitive with some current methods, such as the 
Transfer-Matrix Monte Carlo
method~\cite{Nightingale96}.  

\begin{figure}
      \epsfxsize=\textwidth
      \epsfbox{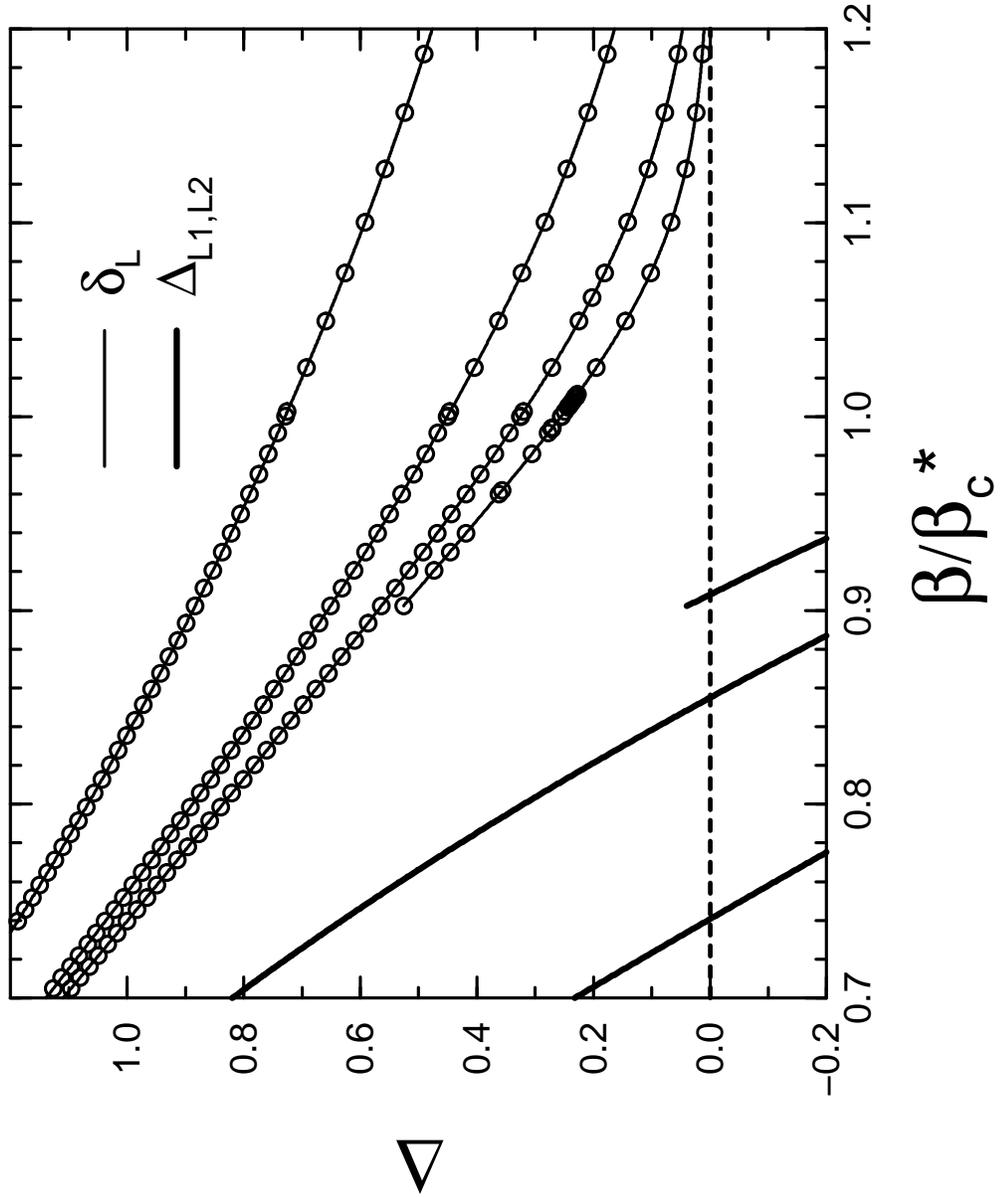}
          \noindent 
        \caption
                { 
                 \label{f:3DGAP}
                 The mass gap for the cubic-lattice Ising ferromagnet. 
                 The thin solid curves are clamped splines to 
                 the transfer-matrix calculations (open circles) of 
                 gaps for semi-infinite
                 systems of cross section $L \! = \! 2$, 3, 4, 
                 and 5, from top to bottom.  
                 The thick solid curves are extrapolations 
                 using \protect\eref{eq:XSolve2}; from 
                 left to right, they represent 
                 $(L_1,L_2)$ = (2,3), (3,4), and (4,5).  
                }
\end{figure}

The method here is basically the same as in the 
previous section, although we are much more restricted 
both in the system sizes available and, for 
$L \! = \! 5$, the number of data available. 
Since references~\cite{MRS0,Richards93} dealt with 
phenomena for $\beta \! > \! \beta_{\rm c}^*$, we have 
only a few points $\delta_{5}(\beta)$ 
with $\beta \! < \! \beta_{\rm c}^*$ [see \fref{f:3DGAP}].  
We use cubic splines to evaluate both $\delta_L(\beta)$ and the first 
$\d \delta_L / \d \beta$ as continuous functions.  
Except at the low-$\beta$ end of 
the $L \! = \! 5$ data, we always use ``natural'' 
boundary conditions on our cubic splines, i.e.\ 
specifying that $\d^2 \delta_L / \d \beta^2 \! = \! 0$ at 
the end points of our data.  For the low-$\beta$ end of 
the $\delta_{5}$ spline, 
we use both natural boundary conditions and ``clamped'' 
boundary conditions by extrapolating 
$\d \delta_{5} / \d \beta$ from the smaller system sizes. 
There is little difference in $\delta_{5}$ itself for these 
two splines. 

We extrapolate $\Delta_{L,L^\prime}$ 
using \eref{eq:XSolve2} and use \eref{eq:hypo}, 
\eref{eq:defY}, and \eref{eq:defX} to 
estimate $\nu$.  As \fref{f:3Dnu} shows, the splines 
yield unreliable estimates near the end of the $\delta_{5}$ 
data. However, as long as $\beta_{\{4,5\},\rm c}$ is reasonably 
large, the estimates of $\nu$ do not depend on the boundary 
conditions used for the spline and seem more reliable. 
Choosing five such points and using Aitken's 
$\Delta^2$ method~\cite{BurdenFaires}~\footnote{The symbol $\Delta^2$ 
is a part of the name of the numerical method and should not be confused 
with either $\Delta(\beta)$ or $\Delta_{\{i\}}(\beta)$.} 
{}to accelerate the convergence, we 
extrapolate to $[\nu(2,3;3,4)-\nu(3,4;4,5)]^{-1} \rightarrow 
\infty$, yielding $\nu \! = \! 0.629$.  Given the very 
small systems used in this estimate, this is in good  
agreement with other recent estimates of $\nu$. 

\section{The Spin-1 Antiferromagnetic Heisenberg Chain}
\label{sec-1DH} 

The Heisenberg chain we study is defined by the quantum Hamiltonian 
\begin{equation}
  \label{eq:HHeisen}
  {\cal H} = \sum_{i=1}^{L} \left[ 
       \vec{S}_i \cdot \vec{S}_{i+1} 
       - \beta (\vec{S}_i \cdot \vec{S}_{i+1})^2 \right] \; ,
\end  {equation}
where $\vec{S}_i$ is the quantum spin-1 operator for the spin at 
site $i$ and $\vec{S}_{L+1} \! \equiv \! \vec{S}_1$ 
(periodic boundary conditions). 
In the limit $L \! \rightarrow \! \infty$,
the spectrum of this Hamiltonian has been exactly 
solved for at $\beta \! = \! 1$
using Bethe-ansatz techniques, and the resulting energy spectrum 
is gapless~\cite{Takhtajan82,Babujian83a,Babujian83b}.  
The ground state of the Hamiltonian 
has also been found~\cite{Affleck87, Affleck88} for even 
values of $L$ at $\beta \! = \! -1/3$, 
and the gap above it has been proven to be nonvanishing.
It has been argued~\cite{AffleckNP86} 
that $\beta \! = \! 1$ marks a 
phase transition between the gapped Haldane 
phase~\cite{Haldane83a,Haldane83b} 
and the gapped dimerized phase, and that this 
phase transition is in the same universality class as 
the two-dimensional Ising model.  These arguments have been 
supported numerically~\cite{Bloete86,Fath93}.

\begin{figure}
      \epsfxsize=\textwidth
      \epsfbox{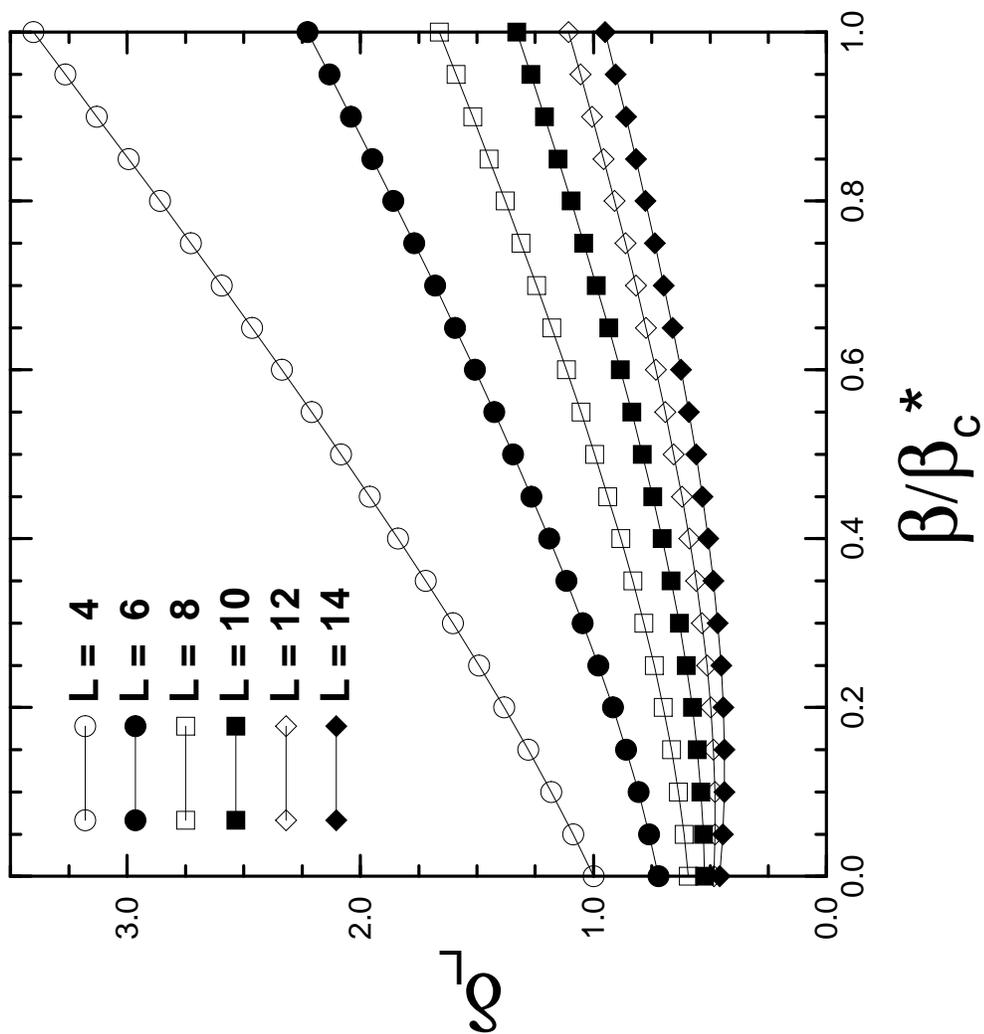}
          \noindent 
        \caption
                { 
                 \label{f:S1gapL}
                 Gap estimates $\delta_L$ for the 
                 spin-1 Heisenberg chain with periodic 
                 boundary conditions.  Strong finite-size 
                 effects obscure the fact that the gap 
                 vanishes at $\beta \! = \! 1$ in the 
                 limit $L \! \rightarrow \! \infty$.
                }
\end{figure}

\begin{figure}
      \epsfxsize=\textwidth
      \epsfbox{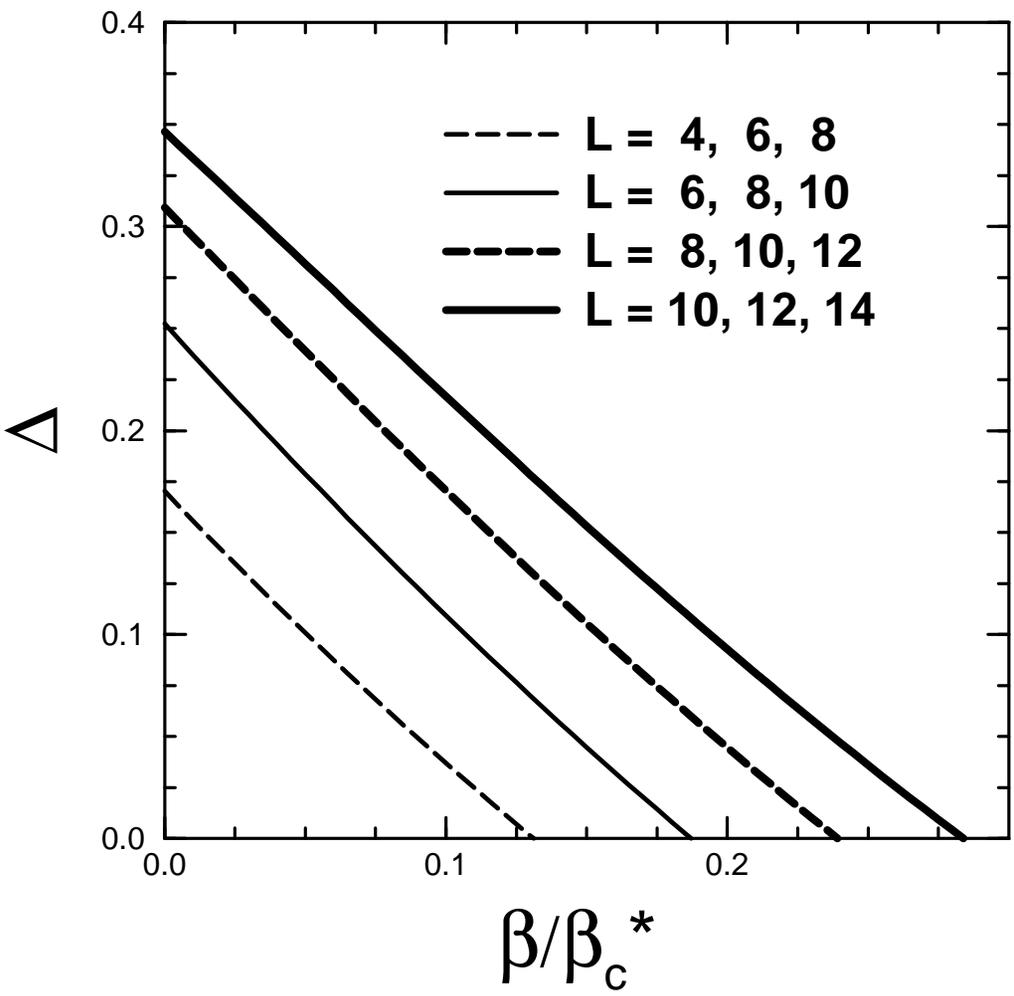}
          \noindent 
        \caption
                { 
                 \label{f:S1gapE}
                 Extrapolated gap estimates $\Delta_{\{L\}}$ for the 
                 spin-1 Heisenberg chain with periodic 
                 boundary conditions.  
                }
\end{figure}

\Fref{f:S1gapL} shows the estimates $\delta_L(\beta)$ obtained 
by exact diagonalization of small systems.  The finite-size 
effects are quite strong, and we were unable to make an 
extrapolation of the form \eref{eq:XSolve2}, since we require 
$0 \! \leq \! \beta_{\{L\},\rm c} \! < \! 1$ for all 
extrapolations in order for $X$ and $Y$ to be real and finite. 
Instead we solve 
\begin{equation}
  \eqalign{
  \label{eq:XSolve3a} 
  \delta_{L}  (\beta) &= D(\beta) + A(\beta) L^{  -B(\beta)} \\
  \delta_{L'} (\beta) &= D(\beta) + A(\beta) L'^{ -B(\beta)} \\
  \delta_{L''}(\beta) &= D(\beta) + A(\beta) L''^{-B(\beta)} 
          }
\end  {equation} 
for $B(\beta)$ at fixed $\beta$, 
and then solve 
\begin{equation}
  \eqalign{
  \label{eq:XSolve3b} 
  \delta_{L'} (\beta) &= \Delta_{L,L',L''}(\beta) 
            + C(\beta) L'^{ -ZB(\beta)} \\
  \delta_{L''}(\beta) &= \Delta_{L,L',L''}(\beta) 
            + C(\beta)L''^{-ZB(\beta)} 
          } 
\end  {equation} 
for $\Delta_{L,L',L''}(\beta)$ using the value of 
$B(\beta)$ found from \eref{eq:XSolve3a}.  
Here $L \! < \! L'\! < \! L''$. 
The parameter $Z$ allows us to 
tune the extrapolation as in the preceding sections. 
\Fref{f:S1gapE} shows an example of extrapolations which 
in fact yield the best estimate of $\nu$. 

\begin{figure}
      \epsfxsize=\textwidth
      \epsfbox{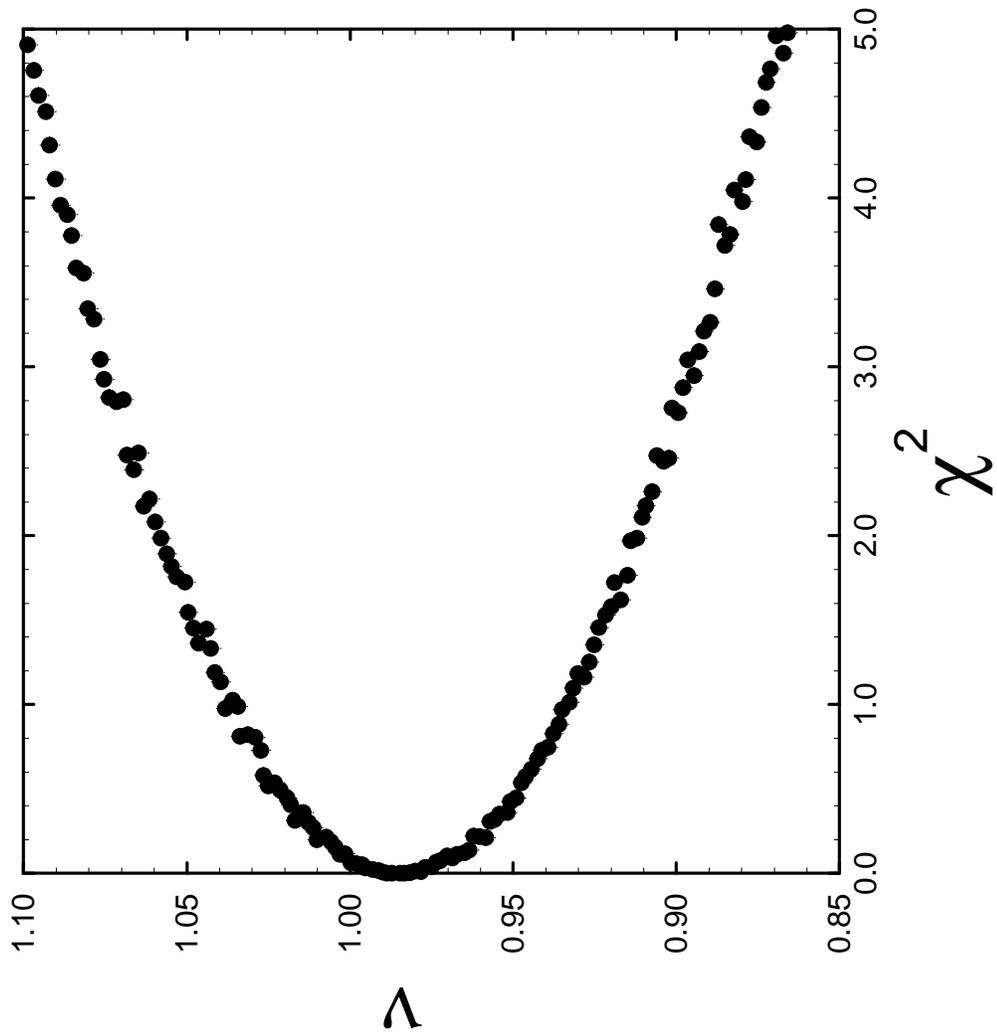}
          \noindent 
        \caption
                { 
                 \label{f:S1nu}
                 We fit the CAM data to equation 
                 \protect\eref{eq:S1fit} while varying 
                 $Z$ in \protect\eref{eq:XSolve3b}. 
                 The CAM data are equally weighted with 
                 an arbitrary fixed weight. 
                 The minimum of $\chi^2$ gives 
                 the best estimate for $\nu$. 
                }
\end{figure}

\begin{figure}
      \epsfxsize=\textwidth
      \epsfbox{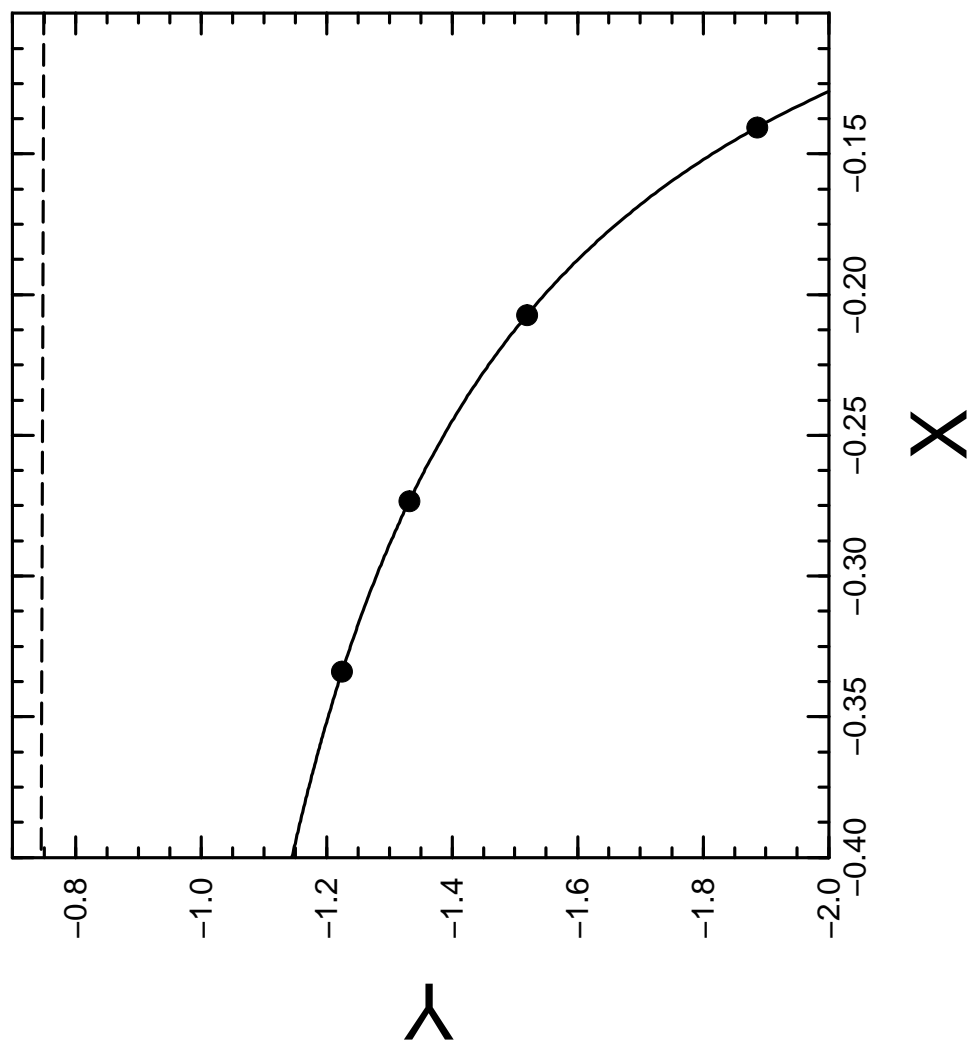}
          \noindent 
        \caption
                { 
                 \label{f:S1CAM}
                 CAM plot for the spin-1 Heisenberg chain. 
                 $Y$ and $X$ are give by 
                 \protect\eref{eq:defY} and 
                 \protect\eref{eq:defX}, respectively. 
                 The solid curve is a fit to 
                 \protect\eref{eq:S1fit}.  The dashed 
                 line is the asymptote of the fitted curve. 
                }
\end{figure}

Even with the extrapolation procedure outlined above, we 
are not able to eliminate the curvature from the CAM plot 
as we did in sections~\ref{sec-2DI} and \ref{sec-3DI}. 
Instead, we perform a fit to the form  
\begin{equation}
  \label{eq:S1fit}
  Y = \frac{a}{X} + b + (\nu - 1) X 
\end  {equation}
while varying $Z$ to find $\nu$, assigning an arbitrary 
fixed weight equally to all of the CAM points. 
The best estimate for $\nu$ is the one that 
minimizes $\chi^2$ [\fref{f:S1nu}]. 
The best fit, shown in \fref{f:S1CAM}, is for 
$\nu \! \approx \! 0.987$, in good agreement with 
theoretical predictions. 

\section{Conclusion}
\label{sec-concl} 

In this paper we propose a new and very general 
method for constructing canonical sequences for use in 
the Coherent Anomaly Method.  
A critical point $\beta \! = \! \beta_{\rm c}^*$
is always marked by the vanishing of 
some quantity $\Delta(\beta)$, though approximations
of $\Delta(\beta)$ often remain nonzero for all 
values of $\beta$. By intentionally 
underestimating the rate of convergence of an initial 
sequence of approximations $\{\delta_i(\beta)\}$, we form extrapolations 
$\{\Delta_{\{i\}}(\beta)\}$ that cross zero at some value of 
$\beta$, which serves as the approximate critical point
$\{\beta_{\{i\},\rm c}\}$.  
Furthermore, $\Delta_{\{i\}}(\beta)$ can very 
generally be expected to cross zero linearly with $\beta$. 
Because there are many ways in which the extrapolations 
can be made from an initial sequence $\{\delta_i\}$, we 
have a great deal of freedom to choose an extrapolation 
which shows a clear coherent anomaly. 

We apply this method to the square lattice Ising model, 
using the exact values of the mass gaps for semi-infinite 
systems of finite width $N$ at temperature 
$T \! = \! \beta^{-1}$
as $\delta_L(\beta)$. 
We find that the method yields $\nu_{L} \! \approx \! 1$ for 
moderate values of $L$ and that 
$\lim_{L \rightarrow \infty} \nu_{L} \! \rightarrow \! 1$, 
as should be expected from the exact solution of the 
two-dimensional Ising model~\cite{Onsager44}. 
The convergence may be complicated and nonmonotonic, however, 
depending on which sets of $L$ are chosen to form the 
extrapolations. 

In order to investigate the convergence of $\nu_{L}$ to $\nu$ 
further, we assume that $\delta_{L}$ follows the scaling 
equation \eref{eq:scale}.  This allows us to show that 
$\nu_{L} \! \simeq \! \nu + O(L^{\phi})$, which is the 
same convergence rate as has been found for finite-size 
scaling~\cite{Privman83}. 

We also apply this method to the cubic-lattice Ising model, 
using numerical transfer-matrix values of the mass gaps 
for semi-infinite systems with $L \! \times \! L$ cross sections 
and periodic boundary conditions.  Even though we are limited 
to systems with $L \! \leq \! 5$ and have a limited amount of 
data for $L \! = \! 5$, we are able to estimate 
$\nu \! \approx \! 0.629$, which is within a few percent of 
the best current estimates of $\nu$~\cite{MCRGTc,Kolesik95,JYChoi97}.  

Finally, we apply this method to a one-dimensional 
spin-1 quantum Heisenberg antiferromagnet, 
using numerical exact-diagonalization estimates of the 
energy gaps for systems of width $L$ and periodic boundary 
conditions.  In spite of large finite-size effects, we are able 
to estimate $\nu \! \approx \! 0.987$, in good agreement with 
theoretical and numerical studies indicating  
$\nu \! = \! 1$~\cite{AffleckNP86,Bloete86,Fath93}. 

\begin{figure}
      \epsfxsize=\textwidth
      \epsfbox{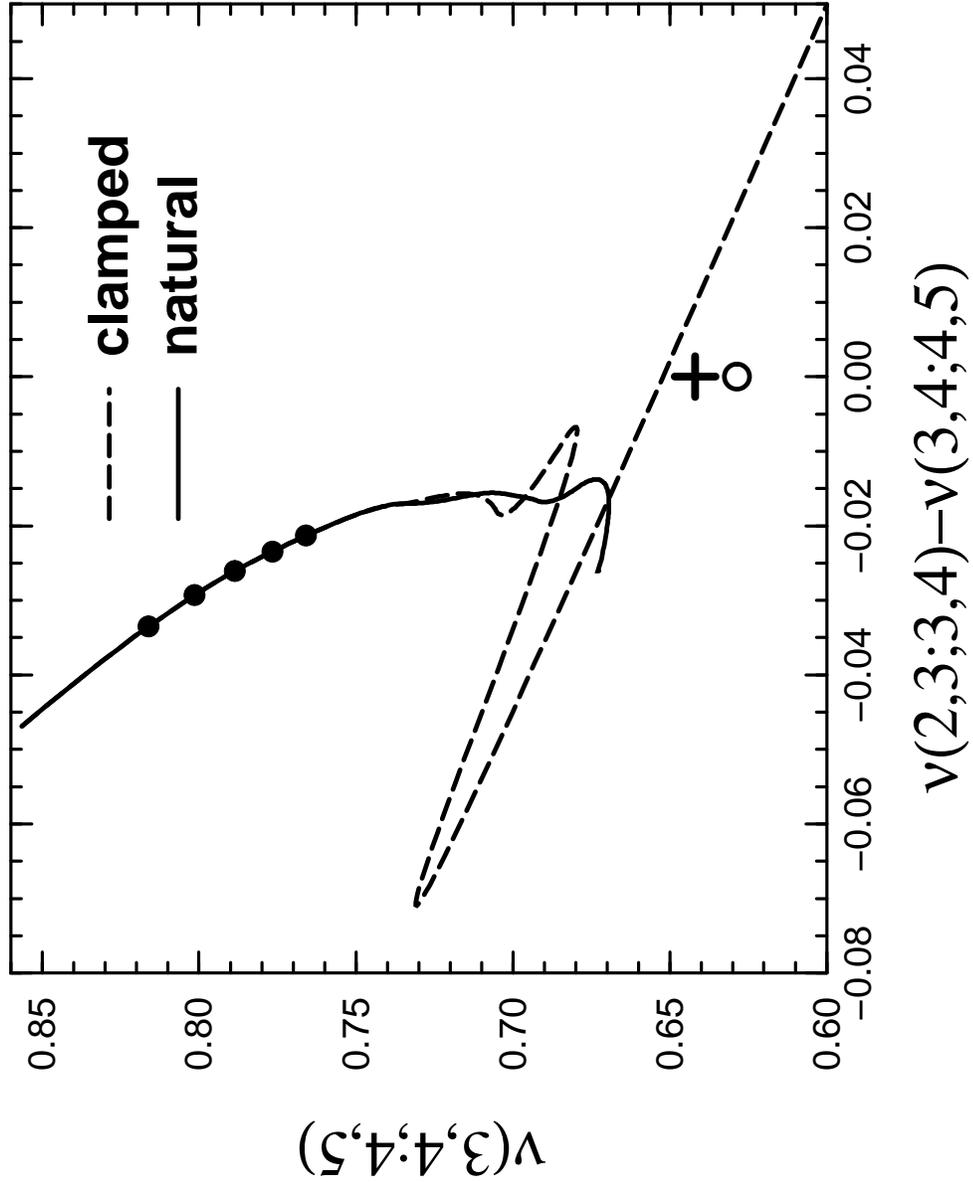}
          \noindent 
        \caption
                { 
                 \label{f:3Dnu}
                 As in \protect\fref{f:2Dnu}, we vary $B$ 
                 to obtain a good estimate of $\nu$. 
                 Results are shown using two different cubic 
                 splines through the data for $L \! = \! 5$. 
                 These curves become meaningless as 
                 $\beta_{\rm c}$ approaches the end of the 
                 data available for $L \! = \! 5$. 
                 By taking five points (solid circles) 
                 where the two curves agree and extrapolating, we 
                 find $\nu \! = \! 0.6285$ (open circle). 
                 This compares well with other estimates, 
                 such as that of \protect\cite{MCRGTc} (cross). 
                }
\end{figure}

There are a few points that need to be emphasized.
\begin{itemize}
  \item Although we use data from finite systems of 
    successively larger size, we are not performing 
    Finite-Size Scaling~\cite{Barber_PTCPv8}.  
    The  initial sequence 
    $\{\delta_i\}$ could be derived from other techniques, 
    such as the Density-Matrix Renormalisation 
    Group algorithm~\cite{White92,White93}, in which 
    the system size is not the most important parameter 
    affecting the quality of the approximation. 
  \item Although we use extrapolations, we are not 
    seeking the {\em best} extrapolations in the 
    sense of extrapolations which are nearest to 
    the thermodynamic limit of the model in question. 
    This is because the extrapolation is just one step 
    in the process.  Instead, we seek a sequence of 
    extrapolations for which the coherent anomaly 
    is clear. 
  \item Taylor expansions have been used previously to form 
    sequences of approximations with 
    $\nu_{\{i\}} \! = \! 1$ from series expansions~\cite{Kinosita92}, 
    but in those studies the linear behaviour of the CAM plot 
    could not be improved, since the physics of the series expansion 
    left no room for change.  The extrapolation-CAM method has 
    flexibility to choose the ``straightest'' CAM plot. 
  \item This method requires good numerical precision. 
    This is clear from \fref{f:3Dnu}.  Care should therefore 
    be exercised when applying this method to Monte Carlo 
    data, where statistical uncertainty may be significant. 
  \item Although we here are estimating only $\nu$, other 
    critical exponents could be found in the same way. 
    For instance, instead of extrapolating the gap, 
    one could extrapolate the reciprocal of the specific 
    heat. 
  \item Some initial sequences $\{\delta_i(\beta)\}$ may already 
    cross zero as $\beta$ is varied.  
    For instance, the variational method 
    of references~\cite{Ostlund95,Rommer97} produces 
    gap estimates for the spin-1 Heisenberg chain 
    that have this property.  In such a case, extrapolations 
    could still be used to look for a clearer coherent 
    anomaly.  The extrapolations should then be faster than 
    the convergence of $\{\delta_i(\beta)\}$, 
    rather than slower, so that 
    the sequence $\{\Delta_{\{i\}}\}$ still crosses zero. 
\end  {itemize}

\ack 

This work was supported by the Inoue Foundation for Science. 
The authors wish to thank Professors M~Kaburagi and H~Nishimori 
for useful comments and for the use of their programs 
{\sc KOBEPACK} and {\sc TITPACK}, respectively.  
The data for the three-dimensional Ising model were 
taken at Florida State University; consult 
reference~\cite{Richards93} for details. 
The authors thank Professor 
Per~Arne Rikvold for comments on the manuscript. 


\bibliographystyle{unsrt}


\end{document}